\RequirePackage{fix-cm}
\documentclass[smallextended,refree]{svjour3}       
\smartqed  
\usepackage{amssymb}
\usepackage{latexsym}
\usepackage[dvipdfmx]{graphicx}
\begin{document}

\title{
Multicritical behavior of the fidelity susceptibility
for the 2D quantum
transverse-field $XY$ model 
}
\subtitle{}


\author{Yoshihiro Nishiyama
}

\institute{Department of Physics, Faculty of Science,
Okayama University, Okayama 700-8530, Japan}

\date{Received: date / Accepted: date}

\maketitle

\begin{abstract}

The two-dimensional  quantum $XY$ model
with a transverse magnetic field
was investigated with the exact diagonalization method.
Upon turning on the magnetic field $h$ and
the $XY$-plane 
anisotropy $\eta$,
there appear a variety of phase boundaries,
which meet at the multicritical point $(h,\eta)=(2,0)$.
We devote ourselves to the Ising-universality
branch, placing an emphasis on the multicritical behavior.
As a probe to detect the underlying phase transitions,
we adopt the fidelity susceptibility $\chi_F$.
The fidelity susceptibility does not rely
on any presumptions as to the order parameter involved.
We made
a finite-size-scaling analysis of $\chi_F$ for
$\eta=1$ (Ising limit), where a number of preceding results are available.
Thereby,
similar analyses with $\eta$ scaled were carried out around the
multicritical point.
We found that 
 the $\chi_F$ data are described by
the crossover scaling theory.
A comparison with the preceding studies of the multicriticality is made.
\end{abstract}



\section{\label{section1}Introduction}

For the quantum mechanical systems,
the fidelity $F$ is defined by the
overlap 
\begin{equation}
F = | \langle h | h + \Delta h \rangle |
,
\end{equation}
between the
ground states with the proximate interaction parameters,
$h$ and $h+\Delta h$;
here, the symbol $| h \rangle$ denotes the ground-state vector
for the Hamiltonian with the interaction parameter $h$.
The idea of fidelity was developed in the course of the study 
of the
quantum dynamics \cite{Uhlmann76,Jozsa94,Peres84,Gorin06}.
Meanwhile,
it turned out that the fidelity is sensitive to the quantum phase transitions.
Actually,
the fidelity susceptibility
\begin{equation}
\label{fidelity_susceptibility}
\chi_F=\frac{1}{N}
\partial_{\Delta h}^2 F|_{\Delta h=0}
, 
\end{equation}
with the system size $N$ 
exhibits a notable peak around the phase transition point
\cite{Quan06,Zanardi06,Zhou08,You11,Rossini18};
see Refs. \cite{Vieira10,Gu10} for a review.
In fact,
 the fidelity susceptibility displays a
pronounced peak as compared to that of  the specific heat
\cite{Albuquerque10}.
As would be apparent from the definition
(\ref{fidelity_susceptibility}),
the fidelity susceptibility is readily tractable 
with the exact diagonalization method.
It has to be mentioned that
the fidelity susceptibility is accessible via the quantum Monte Carlo method
\cite{Albuquerque10,Schwandt09,Grandi11,Wang15}
and the experimental observation \cite{Zhang08,Kolodrubetz13,Gu14}
as well.

In this paper, 
we investigate the two-dimensional quantum $XY$ model
with a transverse magnetic field.
Upon turning on the magnetic field $h$ and the $XY$-plane anisotropy $\eta$,
there appear
a variety of phase boundaries,
which merge at the multicritical point $(h,\eta)=(2,0)$.
We devote ourselves to the Ising-universality branch,
aiming to reveal how the Ising-universality branch ends up at
the $XX$-symmetric multicritical point.
As a probe to detect the phase transitions,
we adopt the fidelity susceptibility.
The fidelity susceptibility is sensitive to both
Ising- and $XX$-symmetric phase transitions,
because
it does 
not rely on any presumptions as to the order parameter involved.
As a matter of fact, 
the similar approaches, namely, the fidelity-susceptibility-mediated analyses
\cite{Luo18}
of the multicriticality \cite{Mukherjee11}, have been made for the
one-dimensional counterpart;
see Refs. \cite{Maziero10,Sun14,Karpat14}
for the information-theoretical approaches as well.
In the present paper, stimulated by these recent developments,
we apply the fidelity-susceptibility-mediated scheme
to the case of two dimensions.
Note that the one-dimensional $XY$ model is 
exactly solvable \cite{Katsura62,Barouch70,Suzuki71},
and the concerned singularities have been investigated in depth.
On the contrary,
such sophisticated techniques are not available in two dimensions,
and the detail still remains an open issue.

\begin{figure}
  \includegraphics{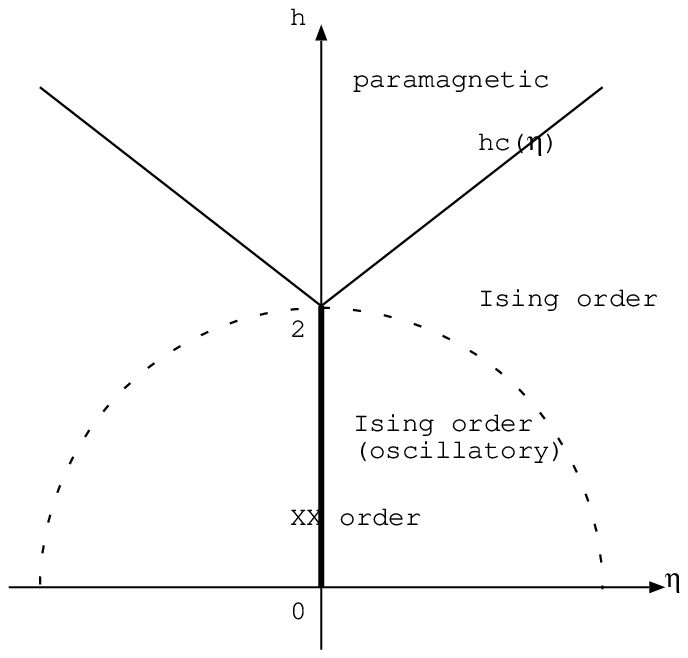}
\caption{
A schematic phase diagram for the two-dimensional
quantum $XY$ model 
with a transverse magnetic field (\ref{Hamiltonian})
is presented.
Here, the parameters, $h$ and $\eta$, denote
the transverse magnetic field and
the $XY$-plane anisotropy, respectively.
For $h > (<) h_c(\eta)$, the paramagnetic
(Ising-ordered) phase extends.
Within the semicircle (dashed), the correlation function gets 
modulated,
and eventually, along the line $\eta=0$
(thick), the $XX$ order is realized.
A topological index 
is specified to each phase surrounding the multicritical point
$(h,\eta)=(2,0)$
\cite{Jalal16},
suggesting a subtlety of this point.
}
\label{figure1}       
\end{figure}

\begin{figure}
  \includegraphics{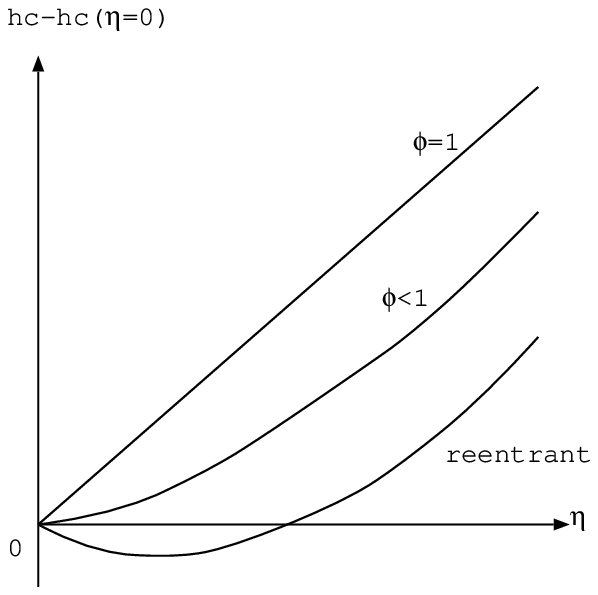}
\caption{
As to the Ising-universality branch $h_c(\eta)$ depicted
in Fig. \ref{figure1},
the spin-anisotropic spherical model displays a reentrant behavior
\cite{Wald15},
whereas the exact diagonalization analysis
for $N \le 5\times 5$
\cite{Henkel84}
suggests
 a ``monotonous'' \cite{Wald15} dependence on $\eta$.
In this paper, based on the crossover scaling analysis
for $N \le 6\times 6$,
we estimate the crossover exponent $\phi$ quantitatively.
}
\label{figure2}       
\end{figure}

To be specific,
we present the Hamiltonian 
for the two-dimensional quantum $XY$ model with a transverse magnetic filed
\cite{Henkel84}
\begin{equation}
\label{Hamiltonian}
{\cal H}=
-J \sum_{\langle ij \rangle} \left((1+\eta)S^x_i S^x_j+(1-\eta)S^y_i S^y_j \right)
-h \sum_{i=1}^{N} S_i^z
.
\end{equation}
Here, the quantum spin-$1/2$
operator ${\bf S}_i$
is placed at each square-lattice point, $i=1,2,\dots,N$
($N=L^2$).
The summation, $\sum_{\langle ij \rangle}$, runs over all possible
nearest-neighbor  pairs, $\langle ij \rangle$;
here,
the periodic boundary condition is imposed.
The coupling $J$ denotes the 
nearest-neighbor ferromagnetic $XY$ interaction,
and it is regarded as the unit of energy
throughout this study;
namely, we set
$J=1$ hereafter.
The parameter $\eta$
denotes the $XY$-plane anisotropy, which interpolates
the Ising ($\eta=1$) and $XX$ ($\eta=0$) symmetric cases smoothly.
The magnetic field $h$ induces the phase transition
between the ordered ($h < h_c(\eta)$)
and disordered ($h > h_c(\eta)$) phases for respective $\eta$.
This phase boundary $h_c(\eta)$
belongs to the Ising universality class \cite{Henkel84}.
A schematic phase diagram \cite{Henkel84} for the model (\ref{Hamiltonian}) 
is presented in Fig. \ref{figure1}.
Within the semicircle (dashed),
the correlation function gets modulated spatially,
and
along the line $\eta=0$ (thick), the $XX$-ordered phase is realized eventually.
Note that the nature of this $XX$-ordered phase differs 
significantly from that of the one-dimensional counterpart
\cite{Luo18,Mukherjee11,Maziero10,Sun14,Karpat14}.
because the latter corresponds to the case
of
the lower critical dimension (Tomonaga-Luttinger liquid),
and the $XX$ order develops only marginally. 
Noticeably enough,
the phase boundaries in Fig. \ref{figure1}
meet at 
the multicritical point $(h,\eta)=(2,0)$;
actually,
a topological-index is specified \cite{Jalal16} to each regime
surrounding the multicritical point. 
Thereby, there arises a problem
how
the Ising-universality branch $h_c(\eta)$
ends 
up at this $XX$-symmetric point;
see Fig. \ref{figure2}.
The exact diagonalization simulation \cite{Henkel84}
for the $N \le 5 \times 5$ cluster
indicates a ``monotonous'' \cite{Wald15} dependence of $h_c(\eta)$
on $\eta$.
On the contrary,
the spin-anisotropic-spherical-model analysis \cite{Wald15} revealed a reentrant behavior
around the multicritical point, claiming that
the simulation data \cite{Henkel84} are 
``too few and too far apart from a final conclusion''
\cite{Wald15}.
The aim of this paper is to explore the multicriticality
with the crossover scaling analysis of the fidelity susceptibility
for the cluster with $N \le 6 \times 6$ spins
so as to estimate the crossover critical exponent $\phi$ quantitatively.

Upon applying a magnetic field $h$ within
the $XX$-symmetric sector $\eta=0$,
the magnetization saturates at
the multicritical point,
where 
a severe slowing down \cite{Kashurnikov99} affects the efficiency of the quantum Monte Carlo simulations.
This 
point
is characterized by
an enhancement of the dynamical critical exponent $z=2$
\cite{Zapf14},
for which
an asymmetry between the real-space and imaginary-time subspaces
emerges.
Hence, special care has to be taken with 
the ratio between the real-space and imaginary-time system sizes,
$L$ and $\beta$, respectively, 
to carry out the scaling analysis properly.
Because the exact diagonalization method admits us to access the
ground state ($\beta \to \infty$) directly,
it is free from such complications as to the size of $\beta$.
Hence, we are able to concentrate only on the $L$-dependent behaviors
as in the ordinary finite-size-scaling analyses.

We recollect
a number of
related studies 
\cite{Albuquerque10,Henkel84,Jongh98,Yu09,Montakhab10,Huang10,Orrs16}
for the two-dimensional
quantum $XY$ (Ising) model (\ref{Hamiltonian})
in Table \ref{table1}.
As indicated,
in order to detect the Ising-universality phase transitions,
there have been proposed various quantifiers
such as 
energy gap,
fidelity susceptibility,
and 
a number of
variants of entanglement measures \cite{Amico08,Horodecki09}.
As shown in the list,
sophisticated quantifiers other than the energy gap
play a significant role in recent studies.
Correspondingly, a variety of simulation techniques,
such as
exact diagonalization (ED),
density matrix renormalization group (DMRG),
tensor product state (TPS),
quantum Monte Carlo (QMC),
and 
tensor network (TN) methods,
have been employed.
As presented,
the exact diagonalization method is applicable
to various types of quantifiers.
So far,
a limiting case
 $\eta =1$ (Ising limit) has come under thorough investigation
specifically,
and 
the phase transition point $h_c$ at this case $\eta =1$ is 
shown for each study.
As mentioned above,
the  overall features for generic $\eta$
have not been investigated very extensively.

\begin{table}
\caption{
Related studies for the two-dimensional
quantum $XY$ model (\ref{Hamiltonian})
are recollected.
So far, a variety of techniques,
such as 
exact diagonalization (ED),
density matrix renormalization group (DMRG),
tensor product state (TPS),
quantum Monte Carlo (QMC),
and 
tensor network (TN) methods,
have been utilized successfully.
As a probe to detect the underlying phase transition,
there have been proposed a number of quantifiers
such as 
energy gap,
fidelity susceptibility,
and a number of
variants of entanglement measures \cite{Amico08,Horodecki09}.
As indicated, in almost all studies,
the Ising limit $\eta =1$ has been undertaken,
and 
the phase transition point $h_c$ at $\eta =1$ is 
shown for each study.
}
\label{table1}       
\begin{tabular}{llll}
\hline\noalign{\smallskip}
 method & quantifier & $\eta$ range undertaken & $h_c|_{\eta=1}$ \\
\noalign{\smallskip}\hline\noalign{\smallskip}
ED \cite{Henkel84} & energy gap & $\eta \le 1$ &  $3.05(1)$ \\  
DMRG \cite{Jongh98} & energy gap & $\eta=1$ &  $3.046$  \\  
ED \cite{Yu09} & fidelity susceptibility & $\eta=1$ &  $2.95(1)$ \\ 
ED \cite{Montakhab10} & multipartite entanglement & $\eta=1$ & $3.040$ \\ 
TPS \cite{Huang10} & entanglement measures & $\eta=1$ & $3.25$  \\  
QMC \cite{Albuquerque10} & fidelity susceptibility & $\eta=1$ & $3.0442(4)$ \\
TN \cite{Orrs16} & bipartite entanglement per bond & $\eta\le 1$ & $3.26$ \\ 
%
ED (this work) & fidelity susceptibility & $\eta \le 1$ & $3.06(2)$ \\
%
\noalign{\smallskip}\hline
\end{tabular}
\end{table}

The rest of this paper is organized as follows.
In the next section,
we present the numerical results.
The finite-size-scaling analysis is shown
for $\eta=1$, where preceeding results are available.
Then,
similar analyses with $\eta$ scaled 
are made with the
crossover scaling theory.
In Sec. \ref{section3},
we address the summary and discussions.

\section{\label{section2}Numerical results}

In this section, we present the numerical results
for the two-dimensional quantum $XY$
model with a transverse magnetic field, Eq. (\ref{Hamiltonian}).
We employed the exact diagonalization method for the cluster
with $N \le 6 \times 6$ spins.
We dwell on the Ising-universality branch,
placing an emphasis on its multicriticality
at $(h,\eta)=(2,0)$.

As mentioned in Introduction,
the multicriticality for
the one-dimensional counterpart
was studied with the fidelity susceptibility \cite{Mukherjee11}.
In this work \cite{Mukherjee11}, 
the authors took a direct route toward the multicritical point,
setting the parameters
like $h-h_c \sim \eta$.
In this direct approach,
the fidelity susceptibility peak splits into 
a series of sub-peaks,
reflecting the intermittent level crossings 
\cite{Henkel84} along the sector $\eta=0$ ($h<h_c$).
In this paper,
in order to avoid such a peak splitting,
we take a different approach to the multicritical point.
We sweep the magnetic field $h$ with $\eta (\ne 0)$ fixed
to a certain constant value,
and consider
the multicritical singularity as a
limiting case of the ordinary Ising universality class.
To this end,
we first consider the case
$\eta=1$,
where a good deal of preceeding results are available.

\subsection{\label{section2_1}
Scaling behavior for the fidelity susceptibility $\chi_F$
at the Ising limit $\eta=1$}

In this section,
we make a finite-size-scaling analysis of the fidelity susceptibility
for the fixed $\eta=1$ (Ising limit);
this scheme sets a basis for
the subsequent crossover scaling analyses in Sec. \ref{section2_3}.

To begin with,
we recollect
a number of formulas relevant to the present survey.
According to Ref. \cite{Albuquerque10},
the fidelity susceptibility diverges
as 
$\chi_F \sim L^{\alpha_F/\nu}$
at the critical point $h=h_c$,
as the system size $L$ enlarges.
Here, the symbols,
$\alpha_F$ and $\nu$,
denote the critical exponents
for the correlation length and fidelity susceptibility,
respectively;
namely,
the former (latter) diverges
as
$\chi_F \sim |h-h_c|^{- \alpha_F }$
($ \xi \sim |h-h_c|^{- \nu}$)
in the vicinity of 
the critical point for sufficiently large $L$.

\begin{figure}
  \includegraphics{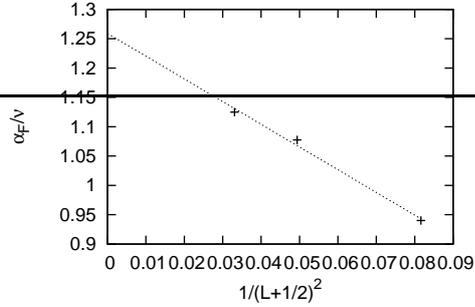}
\caption{The approximate critical exponent
$\frac{\alpha_F}{\nu}(L,L+1)$ 
(\ref{approximate_critical_exponent})
is plotted for $1/(L+\frac{1}{2})^2$
with $\eta=1$ and various system sizes $L=3,4,5$.
The least-squares fit to these data yields
$\alpha_F/\nu=1.259(20)$ in the thermodynamic limit $L\to\infty$.
A possible systematic error is considered in the text.
}
\label{figure3}       
\end{figure}

In Fig. \ref{figure3},
we present the approximate critical exponent
\begin{equation}
\label{approximate_critical_exponent}
\alpha_F /\nu=
\frac{\ln \chi_F(L+1)|_{h=h_c(L+1)} - \ln \chi_F(L)|_{h_c(L)}}
   {\ln (L+1) - \ln L} ,
\end{equation}
as a function of
$1/(L+\frac{1}{2})^2$ with the fixed $\eta=1$ and 
various system sizes $L=3,4,5$;
what is meant by the expression,
$L+\frac{1}{2}$, in the abscissa scale is that the approximate critical exponent,
$\frac{\alpha_F}{\nu}(L,L+1)$ (\ref{approximate_critical_exponent}),
is calculated for a pair of system sizes, $L$ and $L+1$,
and the arithmetic mean, $L+\frac{1}{2}$, is taken as a representative value.
Here, the approximate critical point $h_c(L)$
denotes $\chi_F$'s peak position 
\begin{equation}
\label{approximate_critical_point}
\partial_h \chi_F(L) |_{h=h_c(L)} =0
,
\end{equation}
for each $L$.
The least-squares fit to the data 
in Fig. \ref{figure3}
yields an
estimate
$\alpha_F/\nu=1.259(20)$ 
in the thermodynamic limit $L \to \infty$.
As a reference, we carried out the similar analysis for a pair of data points, $L=4$
and $5$, and arrived at an extrapolated value
 $\alpha_F/\nu = 1.221$.
Regarding the deviation 
 $\approx 0.038$ from the above estimate $1.259$
as a possible systematic error, we estimate the 
 critical exponent as
$\alpha_F/\nu =1.259(38)$.
Putting this estimate into the
scaling relation \cite{Albuquerque10}
\begin{equation}
\label{scaling_relation}
\frac{\alpha_F}{\nu}=\frac{2}{\nu}-2
,
\end{equation}
we arrive at  the correlation-length critical exponent
\begin{equation}
\label{correlation_length_critical_exponent}
\nu=0.614(8) .
\end{equation}
Afterward, we make a comparison with the related studies.

\begin{figure}
  \includegraphics{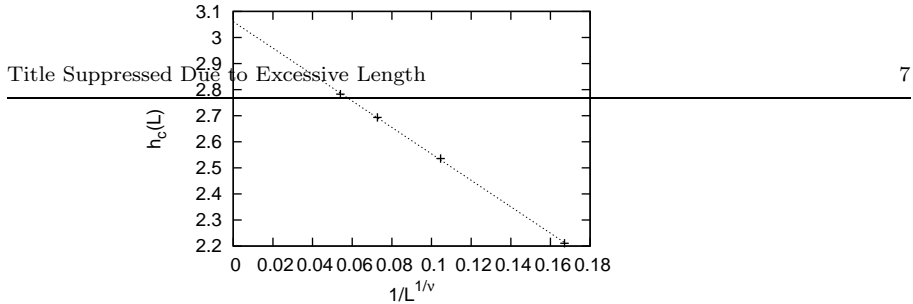}
\caption{
The approximate critical point
$h_c(L)$ (\ref{approximate_critical_point})
is plotted for $1/L^{1/\nu}$
with $\eta=1$, $\nu=0.614$ 
[Eq. (\ref{correlation_length_critical_exponent})],
and various system sizes $L=3,4,\dots,6$.
The least-squares fit to these data
yields an estimate 
$h_c=0.3061(7)$
in the thermodynamic limit $L\to\infty$.
A possible systematic error is considered in the text.
}
\label{figure4}       
\end{figure}

We turn to the analysis of the critical point $h_c$.
The approximate critical point $h_c(L)$ converges
to the thermodynamic limit as $|h_c(L)-h_c| \sim 1/L^{1/\nu} $.
(This relation is anticipated from the above-mentioned formula
$|h-h_c|^{- \nu} \sim \xi $ 
through the dimensional analysis.)
In Fig. \ref{figure4},
we present the approximate critical point
$h_c(L)$ 
as a function of
$1/L^{1/\nu}$ with 
$\nu=0.614$ 
[Eq. (\ref{correlation_length_critical_exponent})],
$\eta=1$,
 and various system sizes 
$L=3,4,\dots,6$.
The least-squares fit to these data yields an estimate 
$h_c=3.061(7)$
in the thermodynamic limit $L \to \infty$.
As a reference, 
we carried out the similar analysis for a pair of 
data points, $L=5$ and $6$,
and arrived at an extrapolated value
 $h_c=3.042$.
Regarding the
deviation $ \approx 0.019$ from the above estimate $3.061$ as a possible systematic error,
we estimate the critical point
as
\begin{equation}
\label{critical_point}
h_c=3.06(2)
.
\end{equation}

This is a good position to address an overview of the related studies
for $\eta=1$.
By means of the 
 fidelity-susceptibility-mediated analyses
with the exact diagonalization 
method for $N \le 20$ 
\cite{Yu09}
and  the quantum Monte Carlo simulation
for $N \le 48^2$
\cite{Albuquerque10},
the estimates,
$(h_c,\nu)=(2.95(1),0.7143)$  
and
$( 3.0442(4) , 0.625(3)  )$,
respectively, 
were obtained.
Our results,
 Eqs.
 (\ref{correlation_length_critical_exponent}) and
(\ref{critical_point}),
are comparable with
 these pioneering studies.
By means of the exact diagonalization method
for $N \le 5^2$
with respect to
the energy gap \cite{Henkel84} and multipartite entanglement
\cite{Montakhab10}, 
there have been reported the results,
$(3.05(1) , 0.629(2))$  
and
$(3.040,0.51)$, respectively.
The former estimates were obtained 
by taking the weighted mean values 
for the simulation results performed at $\eta=0.5,0.7,1.0$
under
the periodic- and anti-periodic-boundary conditions independently.
We stress that the present approach attains 
to admitting rather
 unbiased estimates
with moderate computational effort.
%
Similarly, via the density-matrix-renormalization-group
\cite{Jongh98},
tensor-product-state 
\cite{Huang10},
and
tensor-network
\cite{Orrs16}
methods,       
the estimates,
$(h_c,\nu)=(3.046 , 0.66)$,   
$h_c=3.25$,
and
$3.26$,
respectively, were obtained.
Again, it is suggested that the fidelity-susceptibility-mediated scheme,
albeit with the tractable system sizes restricted,
yields unbiased estimates for the criticality.

\subsection{\label{section2_2}
Scaling plot for the fidelity susceptibility $\chi_F$ at the Ising limit $\eta=1$}

In this section, 
we present the scaling plot for the fidelity susceptibility,
based on the finite-size-scaling formula
\cite{Albuquerque10}
\begin{equation}
\label{finite_size_scaling}
\chi_F=L^x f((h-h_c)L^{1/\nu})
,
\end{equation}
with
the scaling dimension $x=\alpha_F/\nu$
and
a certain (non-universal) scaling function $f$.

\begin{figure}
  \includegraphics{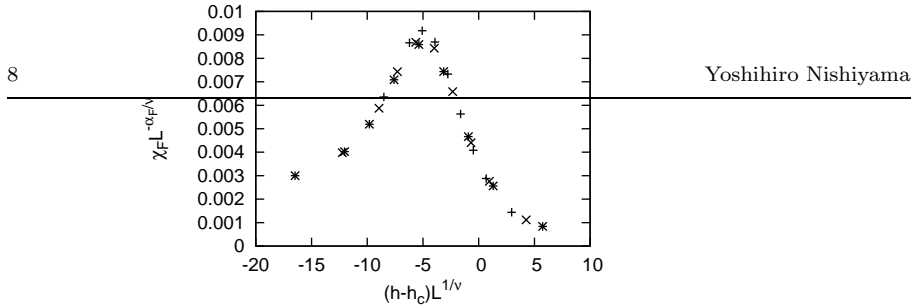}
\caption{
The scaling plot,
$(h-h_c)L^{1/\nu}$-$\chi_F L^{-\alpha_F/\nu}$,
is presented for 
$\eta=1$, 
$h_c=3.06$ [Eq. (\ref{critical_point})],
$\nu=0.614$ [Eq. (\ref{correlation_length_critical_exponent})],
$\alpha_F/\nu=1.259$ [Eq. (\ref{scaling_relation})],
and
various system sizes
($+$) $L=4$,
($\times$) $5$, and
($*$) $6$.
The data collapse into a scaling curve,
validating the finite-size-scaling analyses for the fidelity susceptibility.
}
\label{figure5}       
\end{figure}

In Fig. \ref{figure5},
we present the scaling plot, 
$(h-h_c)L^{1/\nu}$-$\chi_F L^{-\alpha_F/\nu}$,
for $\eta=1$ and various system sizes,
$L=4,5,6$; the symbol for each $L$ is explained in the figure caption.
Here, the scaling parameters,
$h_c=3.06$ [Eq. (\ref{critical_point})],
$\nu=0.614$ [Eq. (\ref{correlation_length_critical_exponent})],
and
$\alpha_F/\nu=1.259$ [Eq. (\ref{scaling_relation})],
were determined in Sec. \ref{section2_1}.
The data in Fig. \ref{figure5} seem to collapse into a scaling curve 
$f$ satisfactorily,
validating the consistency of the analyses
in Sec. \ref{section2_1}.
In the next section, 
with $\eta$ varied,
the data are recast into an extended scaling formula,
namely, the crossover scaling theory,
so as to investigate the multicriticality
at $(h,\eta)=(2,0)$.

Last, we address a number of remarks.
First,
as presented in Fig. \ref{figure5},
the fidelity susceptibility 
exhibits a notable peak around the critical point.
Actually, the scaling dimension 
$\alpha_F/\nu$ of the fidelity susceptibility 
is larger than that of the specific heat
$\alpha/\nu$
because of
the relation
$\alpha_F/\nu=\alpha/\nu+1$ between them.
(This relation is derived from the hyper-scaling relation
$\alpha=2-3\nu$ together with Eq. (\ref{scaling_relation}).)
The fidelity susceptibility 
has an advantage in that its enhanced singularity
would dominate the regular (non-singular) part.
Last,
the fidelity susceptibility does not rely on any
{\it ad hoc} assumptions for the order parameter involved.
Such a feature is significant in the present study,
because 
the crossover between the Ising- and $XX$-symmetric cases
 is our concern,
Rather intriguingly, 
at the $XX$-symmetric point,
the fidelity susceptibility exhibits even more
pronounced peak (larger scaling dimension),
as explained below.

\subsection{\label{section2_3}
Crossover scaling analyses for the fidelity susceptibility 
$\chi_F$ with $\eta$ scaled }

In this section, we carry out the crossover scaling analyses for 
the fidelity susceptibility around $\eta=0$.
Because in this case, an extra parameter $\eta$,
which is supposed to converge to $\eta \to 0$ ($L\to\infty$),
exists,
the above-mentioned scaling formula (\ref{finite_size_scaling})
has to be extended.
According to the crossover scaling theory
\cite{Riedel69,Pfeuty74},
the extended formula should read
\begin{equation}
\label{finite_size_scaling2}
\chi_F = L^{\dot{x}} g
\left(
   \left(h-h_c(\eta)\right) L^{1/\dot{\nu}} ,
   \eta L^{\phi/\dot{\nu}} 
   \right)
,
\end{equation}
with the crossover exponent $\phi$ and a certain 
(non-universal) scaling function $g$;
the crossover exponent $\phi$
\cite{Riedel69,Pfeuty74}
describes how the Ising universality for $\eta>0$
turns into
 the end-point singularity
at $\eta=0$.
As in Eq. (\ref{finite_size_scaling}),
the indices, $\dot{x}=3$ and $\dot{\nu}=1/2$
\cite{Zapf14,Adamski15,Hoeger85},
describe the singularities for
the fidelity susceptibility and correlation length,
 respectively,
right at the multicritical point $\eta=0$.
The former
index
$\dot{x}=3$ is given by
the scaling relation
$\alpha_F/\nu=(\alpha+z\nu)/\nu$ \cite{Albuquerque10}
substituted 
with 
$\nu=1/2$ \cite{Zapf14,Adamski15,Hoeger85},
dynamical critical exponent $z=2$ \cite{Zapf14},
and  specific-heat critical exponent $\alpha = 1/2$;
here, this index $\alpha=1/2$ is read off from the 
first derivative of the 
magnetization $m \propto - \sqrt{h_c-h}$ \cite{Zapf14} 
regarded as the internal energy.

As presented in Fig. \ref{figure2},
the crossover exponent $\phi$ determines the
shape of the phase boundary in the vicinity of the multicritical point.
Note that both arguments of the scaling function $g$ in Eq. (\ref{finite_size_scaling2})
should be dimensionless.
Hence,
the scaling dimensions for
 $h-h_c(\eta)$ and $\eta^{1/\phi}$ are identical,
admitting the relation
$h_c -h_c(0) \sim \eta^{1/\phi}$ \cite{Riedel69,Pfeuty74}.
Therefore, the crossover scaling analysis
has implications for the power-law singularity of the phase boundary.

\begin{figure}
  \includegraphics{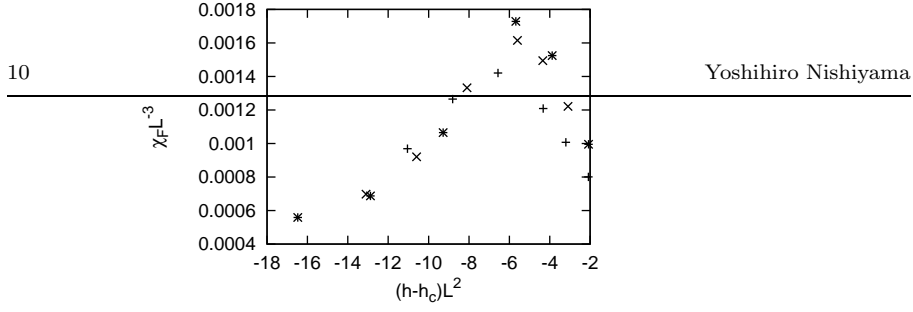}
\caption{
The crossover
scaling plot,
$(h-h_c)L^2$-$\chi_F L^{-3}$,
is presented for 
various system sizes
($+$) $L=4$,
($\times$) $5$, and
($*$) $6$ with 
the fixed 
$\eta L^{2\phi}=10.8$ 
(second argument of the crossover scaling function $g$
(\ref{finite_size_scaling}))
under setting
$\phi=1$.
The crossover-scaled data collapse into a scaling curve,
supporting 
$\phi=1$.
}
\label{figure6}       
\end{figure}

We turn to the crossover-scaling analysis of the fidelity susceptibility,
based on the above-mentioned formulas.
In Fig. \ref{figure6},
we present the crossover scaling plot,
$(h-h_c(\eta))L^2$-$\chi_F L^{-3}$,
for 
various system sizes,
$L=4,5,6$;
the symbol for each $L$ is explained in the figure caption.
Here, the second argument of the scaling function $g$
 (\ref{finite_size_scaling2})
is fixed to a constant value $\eta L^{2\phi}=10.8$
with $\phi=1$,
and the critical point $h_c$ was determined through the same scheme as that of Sec. \ref{section2_1}
by using the index $\nu=0.63002$ \cite{Hasenbusch10}.
The crossover-scaled data in Fig. \ref{figure6} seem to collapse into a scaling curve
satisfactorily.
Actually,
the data for ($\times$) $L=5$ and 
($*$) $6$ almost overlap each other, entering at the 
crossover-scaling regime.
Such a feature strongly supports the proposition
 $\phi=1$.

\begin{figure}
  \includegraphics{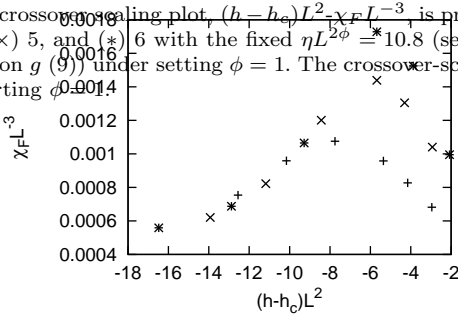}
\caption{
The crossover
scaling plot,
$(h-h_c)L^2$-$\chi_F L^{-3}$,
is presented for 
various system sizes
($+$) $L=4$,
($\times$) $5$, and
($*$) $6$ with 
the fixed 
$\eta L^{2\phi}=26.5$ 
(second argument of the crossover scaling function $g$
(\ref{finite_size_scaling}))
under postulating
$\phi=1.25$.
The right-side-slope 
data get scattered,
as compared to those of Fig. \ref{figure6}.
}
\label{figure7}       
\end{figure}

\begin{figure}
  \includegraphics{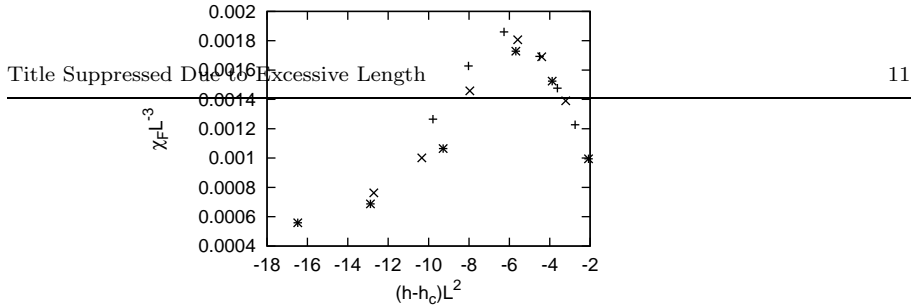}
\caption{
The crossover
scaling plot,
$(h-h_c)L^2$-$\chi_F L^{-3}$,
is presented for 
various system sizes
($+$) $L=4$,
($\times$) $5$, and
($*$) $6$ with 
the fixed 
$\eta L^{2\phi}=4.4$ 
(second argument of the crossover scaling function $g$
(\ref{finite_size_scaling}))
under postulating
$\phi=0.75$.
The left-side-slope data around
$(h-h_c)L^2 \approx -10$ become dissolved,
as compared to those of Fig. \ref{figure6}.
}
\label{figure8}       
\end{figure}

Similar analyses were made for various values of $\phi$.
In Fig. \ref{figure7},
we present the crossover scaling plot,
$(h-h_c(\eta))L^2$-$\chi_F L^{-3}$,
for 
the system sizes, 
$L=4,5,6$,
with the fixed $\eta L^{2\phi}=26.5$ under postulating $\phi=1.25$.
The crossover-scaled data get scattered, as compared to those of Fig. \ref{figure6};
particularly,
the right-side slope displays notable dispersion of the data,
whereas the left hand side shows an alignment.
Additionally,
in Fig. \ref{figure8},
we present the crossover scaling plot,
$(h-h_c(\eta))L^2$-$\chi_F L^{-3}$,
for 
various system sizes, 
$L=4,5,6$,
with fixed $\eta L^{2\phi}=4.4$ under setting $\phi=0.75$.
For such a small $\phi$, on the contrary,
the left-side-slope data around $(h-h_c)L^2 \approx -10$
become dissolved, whereas the right hand side shows a tolerable overlap.
Hence, we conclude that
the crossover critical exponent locates within 
\begin{equation}
\phi=1.0(2) .
\end{equation}
This result indicates that the phase boundary increases, at least, monotonically 
with the anisotropy $\eta$.

A number of remarks are in order.
First,
the fidelity susceptibility exhibits a notable peak around the 
multicritical point.
As noted above,
the scaling dimension $\dot{x}=3$ at the multicritical point
is even larger than  that of the Ising-universality transition
 $x\approx 1.2\dots$.
Hence, the underlying mechanism behind the crossover scaling plot,
Fig. \ref{figure6}, differs from that of the Ising-universality transition,
Fig. \ref{figure5}.
In this sense, 
the overlap of the scaling plot, Fig. \ref{figure6},
is by no means coincidental,
and rather it
requires the exponent $\phi$ to be finely adjusted.
Last,
the fidelity susceptibility is applicable to both
Ising- and $XX$-symmetric cases.
Actually, according to Ref. \cite{Rossini18},
the fidelity susceptibility detects the first-order phase transitions,
and the (putative) critical exponents make sense. 
Our study owes the initial settings for the multicritical indices,
$\dot{x}=3$ and $\dot{\nu}=1/2$, to this recent development \cite{Rossini18}.

\section{\label{section3}Summary and discussions}

The two-dimensional quantum $XY$ model with a transverse magnetic field 
(\ref{Hamiltonian}) was investigated numerically.
We dwell on the Ising-universality branch,
placing an emphasis on the multicriticality
at $(h,\eta)=(2,0)$,
where 
a slowing down \cite{Kashurnikov99}
affects the efficiency of the quantum Monte Carlo simulations.
By means of 
the exact diagonalization method,
we calculated the fidelity susceptibility in order 
to detect the phase transitions.
With fixed $\eta=1$ 
(Ising limit), the finite-size-scaling analysis of the fidelity susceptibility
was made, and the results are in agreement with those of the preceding studies.
Thereby, with $\eta$ scaled, the crossover scaling analysis was made,
and it turned out that the data are cast into the crossover scaling formula
(\ref{finite_size_scaling2})
rather satisfactorily.
As a consequence, we estimate the crossover exponent 
as $\phi=1.0(2)$.
The preceeding 
exact diagonalization analysis \cite{Henkel84}
indicates a ``monotonous'' \cite{Wald15}
dependence of $h_c(\eta)$ on $\eta$.
Supporting this claim,
our result $\phi=1.0(2)$ 
strongly suggests
 a linear increase of $h_c(\eta)$
with $\eta$.

In regard to the reentrant scenario \cite{Wald15}
advocated for the spin-anisotropic spherical model,
it would be
intriguing to investigate the systems with the extended internal symmetries
such as the
spin-$S=1$ $XY$ chain \cite{Hofstetter96},
and the two-band Hubbard model
\cite{Franco18}.
Actually, the latter exhibits a curved
phase boundary
(see Fig. 10 of Ref. \cite{Franco18}) 
reminiscent of
the reentrant scenario.
This problem will be left for the future study.


{\bf Author contribution statement}

Y.N. conceived the presented idea,
and carried out the numerical simulations.
He analyzed the simulation results, and wrote up
the manuscript.

%
%




\end{document}